%% file: chekanov_sergei_deu.tex
\documentclass[twoside]{dis07}
\usepackage[latin1]{inputenc}
\usepackage[dvips]{graphicx,epsfig,color}
\usepackage{wrapfig,rotating}
\usepackage{amssymb,amsmath,array}
\usepackage{cite}
\usepackage{./mcite}
\input ./zeus_def.tex

\begin{document}
\title{First observation of (anti)deuterons in DIS}

\author{S.~Chekanov
%
%
\vspace{.3cm}\\
For the ZEUS Collaboration
\vspace{0.2cm}\\
DESY Laboratory, 22607, Hamburg, Germany. \\
On leave from the HEP division, Argonne National Laboratory, \\
9700 S.Cass Avenue,
Argonne, IL 60439, USA \\
E-mail: chekanov@mail.desy.de
}

%

\maketitle

\begin{abstract}
First observation of (anti)deuterons in 
deep inelastic $ep$ scattering (DIS) measured with the ZEUS detector at HERA 
is reported.
The production rate of deuterons is higher than that of antideuterons.
However, no asymmetry in the production rate of protons and antiprotons
was found.
The (anti)deuteron yield is approximately three orders of magnitude
smaller than that of (anti)protons, which is consistent
with the world measurements.
\end{abstract}

\section{Introduction}

Deuterons ($d$) 
are loosely bound states whose production in high energy collisions
can hardly be accommodated in the standard fragmentation models.
The measurements of such particles may provide an important
information 
on the structure of fragmentation region~\cite{cmodel} and on  
the formation of multiquark states~\cite{karliner-2004-0412}.

In collisions involving elementary particles,
several measurements have been performed in order to understand
the production of $d(\bar{d})$.  
The cross section of $\bar{d}$ in $e^+e^- \to  q\bar{q} $ 
collisions \cite{argus1,*argus2,*opal1,*aleph_deu}
is lower than that measured in hadronic
$\Upsilon (1S)$ and $\Upsilon (2S)$ decays,
and disagrees with the predictions based on the LUND string model~\cite{Gustafson:1993mm}.
The $(\bar{d})$ rate in $e^+e^- \to  q\bar{q} $ is also lower
than that in hadronic ($pA$~\cite{pA1,*pA2,*pA3}, $pp$~\cite{pp1,*pp2,*Abramov:1986ti}) and
in $\gamma p$ collisions at HERA~\cite{h1deuterons}, but
higher than that in nucleus-nucleus ($AA$) collisions
\cite{aa1,*aa2,*star,*Ahle:1998jv,*Bearden:2002ta,*phenix1}.

According to the  coalescence model~\cite{cmodel} developed for heavy-ion collisions,
the $d$  rate is determined by the overlap between the wave
function of a proton, $p$, and a neutron, $n$, 
with the wave function of a $d$. 
The $d$ cross section is the product of 
single-particle cross sections for $p$  and $n$, with a 
coefficient of proportionality, $B_2$, 
reflecting the spatial size of the fragmentation region emitting the particles.

Unlike studies in $AA$, $Ap$ and $pp$, all
previous measurements in collisions involving elementary
particles have been performed for $\bar{d}$, since
the reconstruction of $d$ requires a careful separation of such states from 
particles produced in 
interactions of colliding beams with residual gas in the beam pipe and
secondary interactions of particles on detector material.
The deep inelastic scattering (DIS) events,
which are characterised by the presence of a
scattered electron in the final state,  
provide an ideal environment for studies of $d$, since 
the background from
non-$ep$ interactions is minimal.
In addition, during the 1996-2000 data taking, the ZEUS detector  
had a small amount of inactive material 
between the interaction region and the central tracking detector (CTD). 
This material consisted of the central
beam-pipe and inner wall of the CTD,
with the overall thickness of $2.9$~mm of Al. 
This leads to a small  absorption rate of $\bar{d}$,
as well as to a small contribution from secondary deuterons.

\begin{center}
\vspace{-1.0cm}
\begin{minipage}[c]{0.48\textwidth}
\includegraphics[width=6.5cm,angle=0]{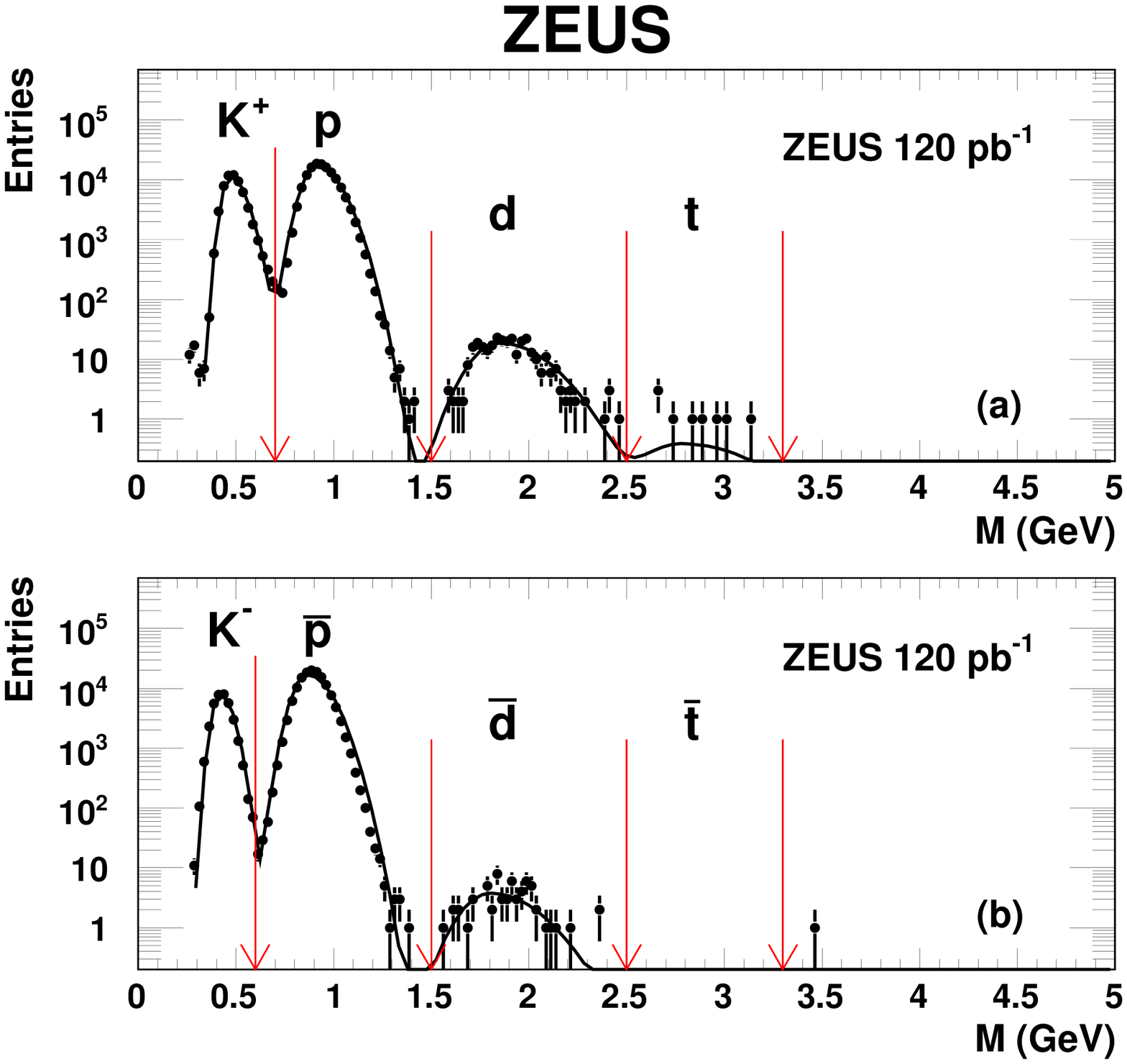}
\label{theta1}
\end{minipage}
\hfill
\begin{minipage}[c]{.48\textwidth}
\includegraphics[width=6.5cm,angle=0]{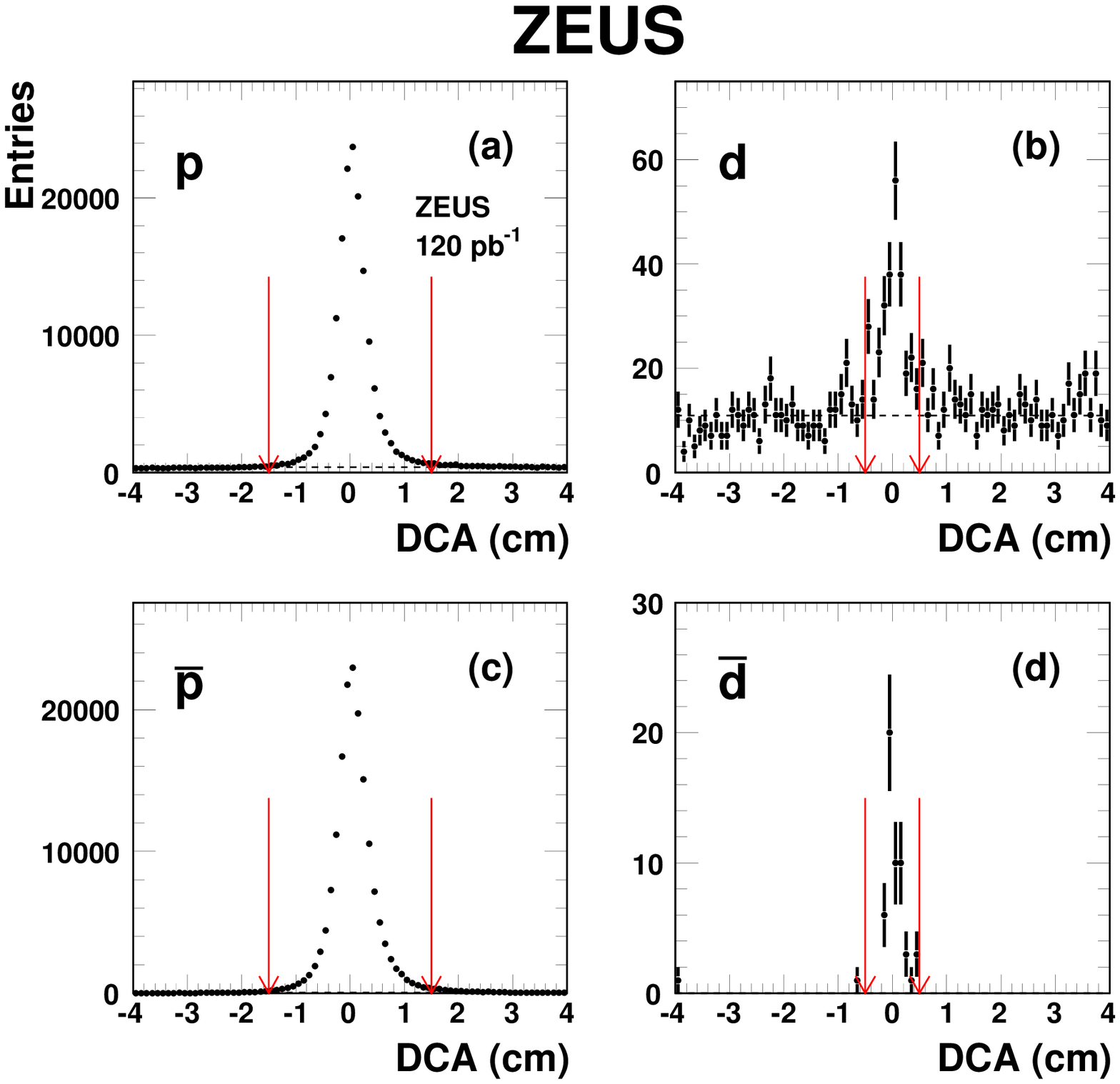}
\label{theta2}
\end{minipage}
\end{center}
\vspace{-0.1cm}
Figure~1. Left: 
The mass spectra reconstructed using $dE/dx$ 
for positive and negative particles.
Right: The $DCA$ distributions 
for: (a)-(b)  particles and  (c)-(d) antiparticles.
\vspace{0.5cm}

The data sample
used in the  analysis corresponds to an integrated luminosity of 120 pb$^{-1}$
taken between 1996 and 2000 with the ZEUS detector at HERA.
After a DIS selection, the average $Q^2$ for the data sample was about 10 GeV$^2$.
The particle identification was performed using the $dE/dx$ measurement.
The masses, shown in Fig.~1(left), were  calculated from the measured track
momentum and energy loss using the Bethe-Bloch formula. 

To identify particles produced in $ep$ collisions,
the distance of closest approach ($DCA$) of the track to the beam 
spot in the transverse plane was used,
since particles originating from the primary $ep$ collisions
feature  small values of $|DCA|$.  
The $DCA$  distribution after the mass
cuts and an additional cut on the distance,  $\Delta Z$,  of the $Z$-component of the
track helix to the primary vertex, is shown in Fig.~1 (right).  
The number of particles produced in $ep$ was
assessed from the $DCA$ distribution by using a side-band
background subtraction.
The numbers  of $d$  and  $\bar{d}$  after the side-band
background subtraction were  $177\pm 17$ and $53\pm 7$, respectively.
This difference was found to be unlikely related to the CTD 
efficiency, which usually leads  to a larger number of negative tracks compared to positive ones: 
For example,
the  number of reconstructed $p$($\bar{p}$) in the data after the $DCA$ side-band
background subtraction
was  $1.52\times 10^5$ ($1.62\times 10^5$). Such $p-\bar{p}$  symmetry is fully accounted for
by known difference in the tracking efficiency for positive and negative tracks.

Several sources of background processes for the $d$ sample  were considered~\cite{2007wz}: 
events due to interactions of the incoming proton (or electron) beam with residual gas in the beam pipe
(termed beam-gas interactions) and 
secondary interactions of particles on material between the interaction point and
the central tracking detector.
Extensive checks have been performed to exclude the first source. In particular,
a special event selection was used for non-colliding electron and proton bunches.
It was found that, after the DIS event selection,  a contribution from the beam-gas events is unlikely.

Even for a clean DIS sample,
$d$ can still be produced by  secondary interactions of particles on
inactive material. One possible source for $d$ is the reaction
$N+N \to  d+\pi^+$, where 
one nucleon, $N$, is produced
by $ep$ collisions, while the second one comes from inactive material
in front of the CTD. Secondary $d$ can also be produced in the pickup reaction  $p+n\to d$ by
primary nucleon interacting in the surrounding material. Checks for such sources of
background were either negative or not conclusive due to insufficient information
on the production cross section of the pickup process. In particular,
a check was done using  the  HERAII data collected  
with the ZEUS detector equipped with a vertex detector.
For the $dE/dx$ measurement,
this additional detector
increased the overall material between the interaction point
and the CTD by a factor three.
As expected, the production rate of $d$  has significantly increased due to a stronger
contribution from spallation processes.  However,  
the rate of $\bar{d}$ after the $DCA$ background subtraction  was the same as for HERAI.

\begin{center}
\vspace{-0.2cm}
\begin{minipage}[c]{0.48\textwidth}
\includegraphics[width=6.6cm,angle=0]{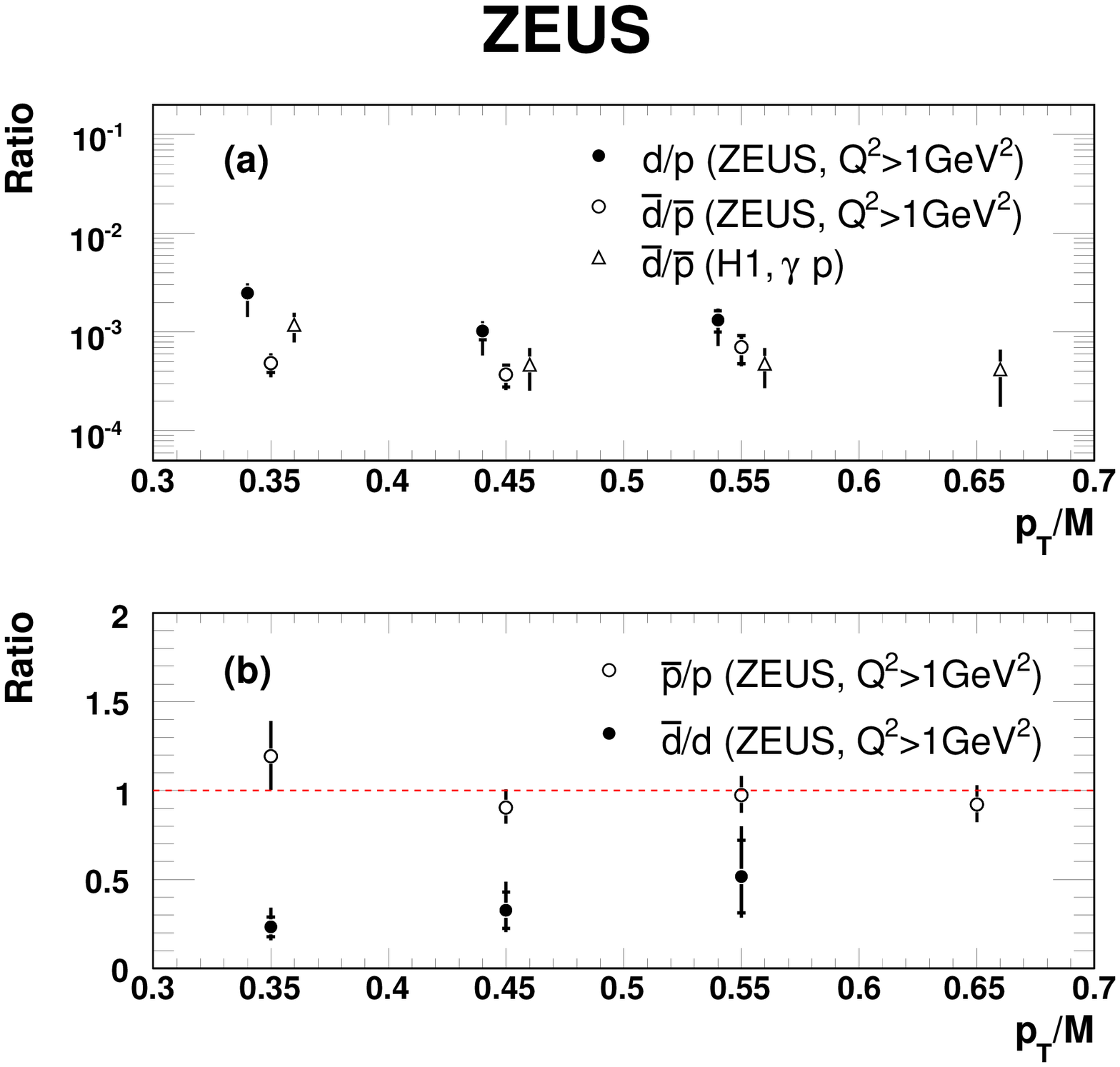}
\label{theta1a}
\end{minipage}
\hfill
\begin{minipage}[c]{.48\textwidth}
 \includegraphics[width=5.5cm,angle=0]{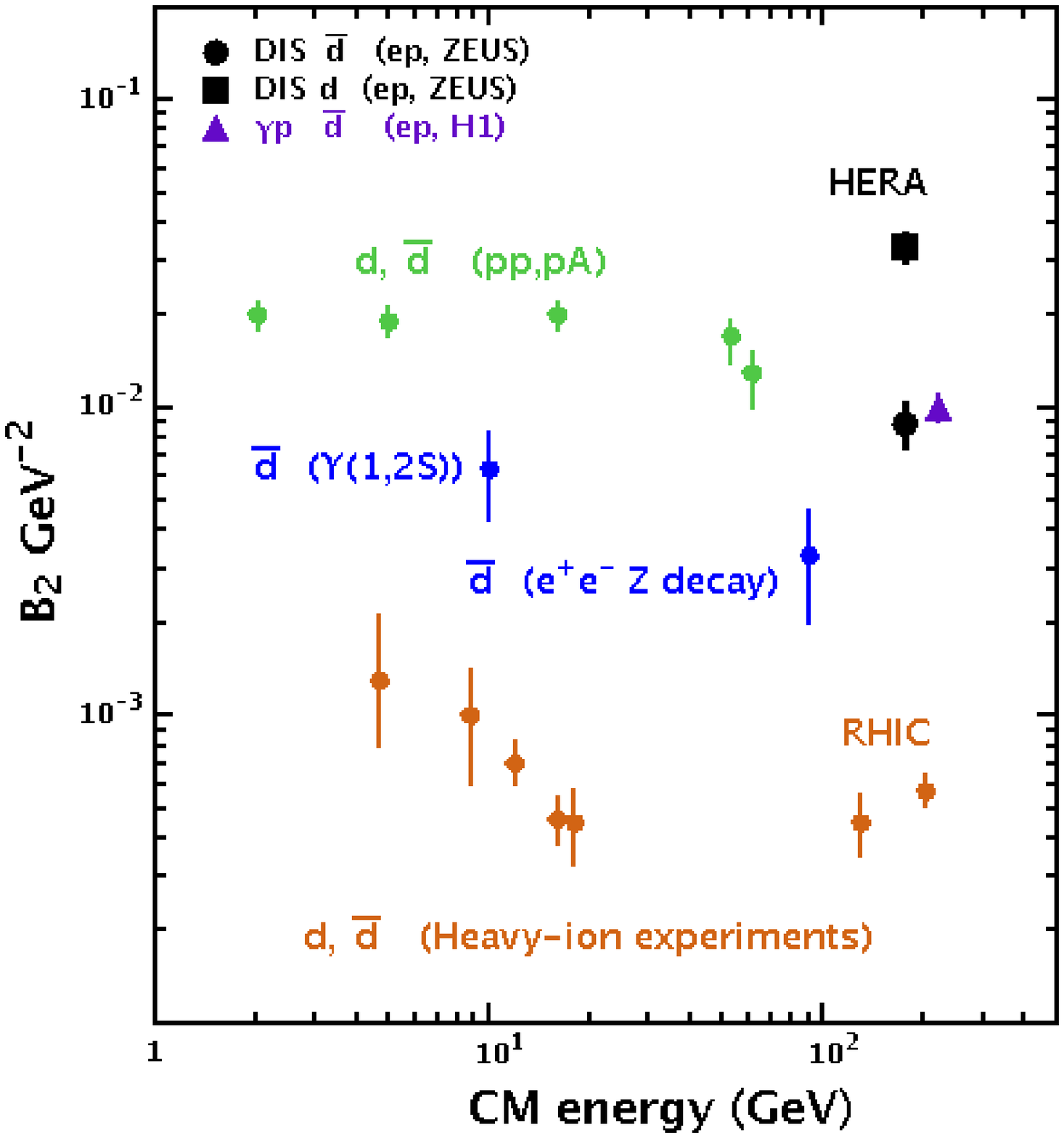}
\label{theta2a}
\end{minipage}
\end{center}
\vspace{-0.5cm}
Figure~2. Left: 
$d/p$,  $\bar{d}/ \bar{p}$, $\bar{p}/ p$ and $\bar{d}/d$ 
production ratios as a function of $p_T/M$. 
Right: Comparison of the  $B_2$ values extracted from DIS with other
world measurements.
\vspace{0.5cm}

The detector-corrected
production ratios as a function of $p_T/M$ are shown in Fig.~2(left).
For the antiparticle ratio,
there is a good agreement with the H1 published data for photoproduction
\cite{h1deuterons}, as well as with $pp$ data \cite{pp1,*pp2,*Abramov:1986ti}.
The production rate of $d$ is higher than that of  $\bar{d}$
at low $p_T$.
If (anti)deuterons are produced as a result of a coalescence of two
(anti)nucleons, then one should expect that the $\bar{d}/d$ ratio
is approximately equal to the $(\bar{p}/p)^2$ ratio for the same $p_T$
per (anti)nucleon, assuming the same source radius for  
particles and antiparticles.
In this case, one should expect $\bar{d}/d \simeq 1$, since the measured
$\bar{p}/p$ ratio is consistent with unity.
Under the assumption that secondary interactions do not produce an
enhancement at $DCA=0$ for the $d$ case, the result would indicate
that the relation between
$\bar{d}/d$ and $(\bar{p}/p)^2$ expected from the coalescence model with the same $B_2$ for
particles and antiparticles does not hold
in the central fragmentation region of $ep$ DIS collisions.
In terms of the coalescence model, the $d$ 
production volume in momentum space is larger 
than that for $\bar{d}$.

For collisions involving incoming baryon beams,  there are
several
models~\cite{Kopeliovich:1996qb,*Kopeliovich:1998ps,*Garvey:2000dk,*Chekanov:2005wf,*Bopp:2004qz,*Bopp:2006bpi}
which
predict $p-\bar{p}$  production asymmetry in the central rapidity region.
The predicted asymmetry can be as high as $7\%$  due to  
the presence of the incoming proton.  
As shown in Fig.~2(left), the experimental data for $p$($\bar{p}$)  are not sufficiently precise
to confirm such expectations. 

There are no predictions for the $d-\bar{d}$ asymmetry.
It is possible that 
theoretical expectations  for
such compound  states are  different than those for the $p-\bar{p}$ asymmetry, 
since $d(\bar{d})$ are not contaminated by a large
contribution 
from the standard baryons produced in quarks and gluon fragmentation.
 
The production of $d(\bar{d})$  was studied in terms of the coalescence model.
The $B_2$ values extracted in the region $0.3<p_T/M<0.7$ are shown in Figure~2(right).
While the measurement of $B_2$  has larger experimental errors than
those in the studies of the production asymmetries, 
it is still seen that $B_2$ for $d$ tends to be higher than that for $\bar{d}$.
The values of $B_2$ for $\bar{d}$
are in agreement with the measurements in photoproduction~\cite{h1deuterons}, but
disagree with the $B_2$ measured in $e^+e^-$ 
annihilation~\cite{argus1,*argus2,*opal1,*aleph_deu} at the $Z$ resonance.
In contrast to
heavy-ion collisions, where  $B_2$ strongly increases 
as a function of $p_T$ due to an expanding source,
the $B_2$ measured in $ep$  does not strongly depend on $p_T/M$~\cite{2007wz}.

In summary,
the first observation of (anti)deuterons in $ep$ collisions in the DIS regime
is presented.
The yield of $d(\bar{d})$ is three orders of magnitude smaller than
that of  $p(\bar{p})$,  which is in broad agreement with other experiments.
The production rate of $p$ is consistent with that of $\bar{p}$, while 
the production rate of $d$ is
higher than that for $\bar{d}$ for the same kinematic region.


\bibliographystyle{./l4z_default}
\def\bibname{\Large\bf References}
\def\refname{\Large\bf References}
\pagestyle{plain}
\bibliography{chekanov_sergei_deu}

\end{document}

%% file: zeus_def.tex
\chardef\usc=95
\chardef\til=126
\catcode`\@=11 
\DeclareRobustCommand\xdotspace{\futurelet\@let@token\@xdotspace}
\def\@xdotspace{%
  \ifx\@let@token.\else
  \ifx\@let@token\bgroup.\else
  \ifx\@let@token\egroup.\else
  \ifx\@let@token\/.\else
  \ifx\@let@token\ .\else
  \ifx\@let@token~.\else
  \ifx\@let@token!.\else
  \ifx\@let@token,.\else
  \ifx\@let@token:.\else
  \ifx\@let@token;.\else
  \ifx\@let@token?.\else
  \ifx\@let@token/.\else
  \ifx\@let@token'.\else
  \ifx\@let@token).\else
  \ifx\@let@token-.\else
  \ifx\@let@token\@xobeysp.\else
  \ifx\@let@token\space.\else
  \ifx\@let@token\@sptoken.\else
   .\space
   \fi\fi\fi\fi\fi\fi\fi\fi\fi\fi\fi\fi\fi\fi\fi\fi\fi\fi}
\catcode`\@=12 

\newcommand{\stru}[2]{%
   \relax\ifmmode\hbox{\vrule height#1 depth#2 width0pt}%
   \else\vrule height#1 depth#2 width0pt\fi}

\newcommand{\Ronum}[1]{\uppercase\expandafter{\romannumeral#1}}
\newcommand{\ronum}[1]{\expandafter{\romannumeral#1}}
\DeclareRobustCommand{\LaTeXZ}{%
  \LaTeX\kern-.05em4\kern-.1em
  {\raisebox{-0.2ex}{$\scriptstyle\text{ZEUS}$}}\xspace}

\DeclareMathAlphabet{\mathbf}{OT1}{cmr}{bx}{sl}
\newcommand{\eVdist}{\kern-0.06667em}

\newcommand{\slashfrac}[2]{%
  \raisebox{0.5ex}{\ensuremath #1}\kern-0.12em/\kern-0.08em
  \raisebox{-.8ex}{\ensuremath #2}}

\newcommand{\sqr}[3]{%
    {\vcenter{\hrule height.#3ex\hbox{\vrule width.#2ex height#1ex
     \kern#1ex\vrule width.#3ex}\hrule height.#2ex}}}

\catcode`\@=11 
\newcommand{\parenbar}{\mathpalette\p@renb@r}
\def\p@renb@r#1#2{\vbox{%
  \ifx#1\scriptscriptstyle \dimen@.7em\dimen@ii.2em\else
  \ifx#1\scriptstyle \dimen@.8em\dimen@ii.25em\else
  \dimen@1em\dimen@ii.4em\fi\fi \offinterlineskip
  \ialign{\hfill##\hfill\cr
    \vbox{\hrule width\dimen@ii}\cr
    \noalign{\vskip-.3ex}%
    \hbox to\dimen@{$\mathchar300\hfil\mathchar301$}\cr
    \noalign{\vskip-.3ex}%
    $#1#2$\cr}}}
\catcode`\@=12 

\newcommand{\IP}{{\rm I$\kern-0.01667em$P}\xspace}

\mathchardef\qsm=63
\mathchardef\pls=43
\mathchardef\mns=512
\mathchardef\plm=518
\mathchardef\eql=61
\mathchardef\smallleft=300
\mathchardef\smallright=301
\mathchardef\les=316
\mathchardef\gre=318
\mathchardef\leq=532
\mathchardef\grq=533

\catcode`\@=11 
\newcounter{pict@width}
\newcounter{pict@height}
\newlength{\pict@scale}
\newcommand{\psfigadd}[4]{%
\setcounter{pict@width}{1*\ratio{#2+\pict@scale/2}{\pict@scale}}
\setcounter{pict@height}{1*\ratio{#3+\pict@scale/2}{\pict@scale}}
\setlength{\unitlength}{\pict@scale}
\hbox to #2{\hspace{-\fill}\begin{picture}(\thepict@width,\thepict@height)
\put(0,0){\psfig{figure=#1,width=#2,height=#3,clip=}}
\SetScale{0.283466457}
\SetWidth{1.763889}
{#4}
\end{picture}}
}
\newcounter{pict@widthfst}
\newcounter{pict@widthscd}
\newcounter{pict@widthtot}
\newcommand{\psfigaddtwo}[7]{%
\setcounter{pict@widthfst}{1*\ratio{#2+\pict@scale/2}{\pict@scale}}
\setcounter{pict@widthscd}{1*\ratio{#2+#4+\pict@scale/2}{\pict@scale}}
\setcounter{pict@widthtot}{1*\ratio{#2+#4+#6+\pict@scale/2}{\pict@scale}}
\setcounter{pict@height}{1*\ratio{#3+\pict@scale/2}{\pict@scale}}
\setlength{\unitlength}{\pict@scale}
\hbox{\hspace{-\fill}\begin{picture}(\thepict@widthtot,\thepict@height)
\put(0,0){\psfig{figure=#1,width=#2,height=#3,clip=}}
\put(\thepict@widthscd,0){\psfig{figure=#5,width=#6,height=#3,clip=}}
\SetScale{0.283466457}
\SetWidth{1.763889}
{#7}
\end{picture}}
}
\newcommand{\psfigror}[4]{%
\setcounter{pict@width}{1*\ratio{#2+\pict@scale/2}{\pict@scale}}
\setcounter{pict@height}{1*\ratio{#3+\pict@scale/2}{\pict@scale}}
\setlength{\unitlength}{\pict@scale}
\hbox{\begin{picture}(\thepict@width,\thepict@height)
\put(0,\thepict@height){\psfig{figure=#1,width=#3,height=#2,clip=,angle=270}}
\SetScale{0.283466457}
\SetWidth{1.763889}
{#4}
\end{picture}}
}
\newcommand{\psfigrol}[4]{%
\setcounter{pict@width}{1*\ratio{#2+\pict@scale/2}{\pict@scale}}
\setcounter{pict@height}{1*\ratio{#3+\pict@scale/2}{\pict@scale}}
\setlength{\unitlength}{\pict@scale}
\hbox{\begin{picture}(\thepict@width,\thepict@height)
\put(0,0){\psfig{figure=#1,width=#3,height=#2,clip=,angle=90}}
\SetScale{0.283466457}
\SetWidth{1.763889}
{#4}
\end{picture}}
}
\catcode`\@=12 
\newlength\listtextwidth

\catcode`\@=11 
\newlength{\@tabfninsert}
\newlength{\@tabfnwidth}
\newcommand{\tabfootnote}[2]{%
  \setlength{\@tabfninsert}{0.8em}
  \setlength{\@tabfnwidth}{\textwidth}
  \addtolength{\@tabfnwidth}{-\@tabfninsert}
  \addtolength{\@tabfnwidth}{-0.4em}
  \noindent\makebox[\@tabfninsert][r]{\footnotesize$^{#1}$\hfil}\hfill%
  \parbox[t]{\@tabfnwidth}{\footnotesize #2\hfill}}
\catcode`\@=12 